\documentclass[preprint,aps]{revtex4}

\pdfoutput=1

\usepackage{amsmath}
\usepackage{graphicx}
\usepackage{bm}
\usepackage{fancyhdr}
\usepackage{amsmath}
\usepackage{graphicx}
\usepackage{amsfonts}
\usepackage{amsbsy}
\usepackage{amssymb}
\usepackage[mathscr]{eucal}

\usepackage{color}

\newcommand{\beq}{\begin{equation}}
\newcommand{\eeq}{\end{equation}}
\newcommand{\ba}{\begin{array}}
\newcommand{\ea}{\end{array}}
\newcommand{\bea}{\begin{eqnarray}}
\newcommand{\eea}{\end{eqnarray}}
\newcommand{\bseq}{\begin{subequations}}
\newcommand{\eseq}{\end{subequations}}

\date{\today}

\begin{document}

\title{Quantum threshold reflection is not a consequence of the badlands region of the potential}
\author{Jakob Petersen and Eli Pollak}
\email{jakob.petersen@weizmann.ac.il,eli.pollak@weizmann.ac.il}
\affiliation{Chemical and Biological Physics Department, Weizmann Institute of Science, 76100, Rehovot,
Israel}
\author{Salvador Miret-Artes}
\email{s.miret@iff.csic.es}
\affiliation{Instituto de F\'{\i}sica Fundamental (CSIC), Serrano 123,
28006 Madrid, Spain }


\begin{abstract}
Quantum threshold reflection is a well known quantum phenomenon which prescribes that at threshold,
except for special circumstances, a quantum particle scattering from any potential, even if attractive
at long range, will be reflected with unit probability.
In the past, this property has been associated with the so-called badlands region of the potential,
where the semiclassical description of the scattering fails due to a rapid spatial variation of the
deBroglie wavelength.
This badlands region occurs far from the strong interaction region of the potential and has therefore
been used to ``explain'' the quantum reflection phenomenon.
In this paper, we show that the badlands region of the interaction potential is immaterial.
The extremely long wavelength of the scattered particle at threshold is much longer than the spatial
extension of the badlands region which therefore does not affect the scattering.
For this purpose, we review the general proof for the existence of quantum threshold reflection to
stress that it is only a consequence of continuity and boundary conditions.
The nonlocal character of the scattering implies that the whole interaction potential is involved in
the phenomenon.
We then provide a detailed numerical study of the threshold scattering of a particle by a Morse potential
especially in the time domain. We compare exact quantum computations with incoherent results obtained from a classical Wigner approximation.
This study shows that close to threshold the time dependent amplitude of the scattered particle is
negligible in the badlands region and that the mean flight time of the particle is not shortened due
to a local reflection from the badlands region.
This study should serve to definitely rule out the badlands region as a qualitative guide to the
properties of quantum threshold reflection.
\end{abstract}


\maketitle


\renewcommand{\theequation}{1.\arabic{equation}} %
\setcounter{section}{0} \setcounter{equation}{0}

\section{Introduction}

In any standard quantum mechanics textbook, one of the first elementary exercises proposed for
one-dimensional potentials is the reflection by a step potential. When the energy of  the incident
particle is greater than the step, the initial wave function is reflected  and transmitted at
the discontinuity point of the potential. At threshold conditions, where the incident energy
approaches the step height, the reflection probability tends to one. This condition is termed total
(quantum) reflection since the transmitted  part of the initial wave function is totally suppressed
and no classical turning point is present. In 1988, Senn \cite{senn} established a theorem for general
one-dimensional potentials showing that the reflection probability goes to unity at threshold
conditions except under the special circumstance that the potential supports a resonance state at threshold.

Threshold conditions and/or laws are of paramount importance in gas-phase collisions and scattering
of atoms by solid surfaces. When the incident energy is close to zero, the corresponding de Broglie
wavelength tends to infinity as the inverse of the square root of the energy and quantum effects are expected to
be critical. In 1936, Lennard-Jones and Devonshire \cite{lennard} first recognized this
behavior in the context of atom-surface interaction.
As far as we know, Kohn \cite{kohn} was the first to show that quantum reflection leads to a zero sticking
probability in threshold particle surface scattering.
As Kohn pointed out, ``it is clearly a quantum interference effect between the incoming and reflected waves.''
Quantum threshold reflection prevents sticking since atoms are not able to come into contact with the surface.
The same phenomenon was reported by C${\hat o}$t\'e et al.\cite{cote} in cold atom collisions although the
authors preferred to name it {\it quantum suppression} since suppression entails some sort of exclusion of
amplitude from certain regions.
These authors also realized that one of the exceptions to the suppression is the presence of a bound state at
threshold with no mention of Senn's work which surprisingly has been overlooked for years in this context.

Most of the theoretical work in this field has been carried out using the semiclassical WKB
framework \cite{mody,trost} since analytical expressions are readily obtained.
Within the primitive WKB theory, reflection is only possible if there exist classical trajectories which
are reflected by the potential.
Therefore, necessarily, quantum reflection, especially for a purely attractive potential cannot be described
by WKB theory (unless one includes higher order corrections or connection formulae).
With this perspective, Friedrich and Trost \cite{trost} conclude that ``Quantum reflection can only occur in
a region of appreciable quantality, i.e. where the condition (36) is violated.''
The condition 36 they mention refers to the badlands function (also known as the quantality function) defined
in Eq.~(\ref{3.20}) below, which is related to the rapidity of the spatial change of the deBroglie wavelength
which must remain small for the WKB approximation to be valid.

At near-threshold conditions, the quantality function becomes very large in a finite spatial region
which is quite far from the surface, where the actual attractive force on the particle is very weak.
Friedrich and coworkers assume that the quantum reflection emanates from this badlands region \cite{trost,friedrich1,friedrich2}.
In Ref.~\cite{friedrich13} Friedrich writes: ``even though there is no potential barrier and no
classical turning point, incoming waves can be partially reflected in this nonclassical region of
coordinate space, so that only a fraction of the incoming radial wave penetrates through to the
near-origin regime. Such classically forbidden reflection ... is called {\it quantum reflection}.''

In Ref.~\cite{friedrich2}, Friedrich and Jurisch
note that ``this nonclassical region is typically located at distances of several hundreds or thousands
of atomic units where the potentials are well described by van der Waals forces with retardation effects.''
In their view, the failure of the WKB approximation in this region provides a local condition for quantum
reflection since it is restricted to occur in a small part of the coordinate space.
This qualitative picture is widely used and accepted.
Stickler et al \cite{stickler15} comment that ``where $B(y)$ is significantly nonzero can be regarded as
regions where quantum reflection can occur'' ($B(y)$ indicates the badlands function).
For example, in their recent measurements on electrically controlled quantum reflection, Barnea et al \cite{barnea17}
note in their Fig.~2 that ``the inset shows the badlands function for different voltages, indicating the region contributing to quantum reflection.''

As mentioned for example by Doak and Chizmeshya \cite{doak00} the WKB approximation breaks down in the
region where the magnitude of the attractive potential energy equals the incident energy.
They then note that ``In a sense, this region can be viewed as an analog at positive kinetic energy to
a classical turning point.''
Mody et al \cite{mody} claim the same, stating that ``The distance away from the slab at which the particle
is turned around -- or quantum reflected -- is precisely this distance.''
Zhang et al \cite{zhang16} point out in the legend to their Fig.~1 which is meant to describe the general
phenomenon of quantum reflection, that ``The quantum reflection probability is non-vanishing in a range of
distances around $z_0$ defined as the distance where the absolute magnitude of the potential energy $|V(z_0)|$
equals the incident kinetic energy $E_z$.''

We do not question the fact that the region in which the WKB approximation breaks down is local, and
reasonably well defined by the region in which the badlands function is greater than unity or that equivalently the
incident kinetic energy is the same as the magnitude of the attractive long range potential.
However, this does not mean that quantum reflection can be described correctly as occurring in this region.
In a recent paper \cite{salva}, the close-coupling formalism (which is numerically exact when convergence
is reached) has been used successfully to describe the experimental work on the quantum reflection of He
atoms from a grating. \cite{zhao1,zhao2}
The interaction potential consisted of two parts, the long range was given by a Casimir-van der Waals (CvW)
tail and the short range by a Morse potential.
In order to avoid the left classical turning point of the repulsive part of the Morse potential which also
leads to reflection and so makes it difficult to distinguish it from the long range quantum reflection, absorbing
boundary conditions were used. \cite{miller,muga}
This was implemented by introducing an imaginary potential which is essentially zero in the physically relevant
interaction region and is turned on at the edge of the coordinate grid for numerical integration, preventing
reflection from the repulsive part of the potential.
The central conclusion of Ref.~\cite{salva} was that quantum reflection is a coherent interference process
which involves the full interaction potential.
It should be considered as a {\it nonlocal effect}.
Furthermore, in this formalism, the dynamics takes place among the different (diffraction) channels needed
for numerical convergence.
The picture that emerged from our previous computation was that at near threshold conditions, quantum reflection
was a nonlocal coherent process.

The purpose of the work presented here is to resolve this issue once and for all.
We will show that the badlands region has nothing to do with quantum reflection and the wavefunction is not
reflected by the badlands region of the attractive potential.
As its name states, quantum reflection is a quantum effect, one proves its existence without the need to resort
to any semiclassical theory.
For this purpose we find it necessary to review in Section 2, Senn's general proof of quantum threshold reflection
in one dimensional systems.
The phenomenon is a direct result of boundary conditions and continuity of the wavefunction and its derivative.
It is thus a global effect and very general.
The potential plays a role in determining how small must the incident kinetic energy be, for quantum reflection to
become important.

We also note that to date, no one has undertaken a quantum time dependent wavepacket propagation to actually
follow the wavepacket in time and see where and whether it is reflected.
This is not an accident, due to the very low energies involved, one needs to evolve a wavepacket which is very
broad in space and moves very slowly.
This is difficult to implement, using numerical wavepacket propagation techniques, even with present day computational resources.
However, if the propagator is known analytically, the problem becomes much easier.

For this purpose, we present in Section~3 a detailed study of the quantum reflection phenomenon for a particle
scattered on a Morse potential.
We derive an explicit expression for the propagator and use it to study the space and time dependence of the
scattering dynamics of the Morse potential at low incident momenta where quantum threshold reflection dominates the dynamics.

Since the tail of the Morse potential is exponential, it has a badlands region and so may be used to study in detail whether
or not the badlands region has any direct effect on the scattering.
We find that indeed, due to the very long wavelength of the incident particle at the ultra low energies leading to quantum
threshold reflection, the large fraction of the density of the particle at all times stays far away from the origin where
the Morse potential has its well.
The lower the energy, the further away the density stays from the origin. However, we show that this distance has nothing to
do with the badlands region of the potential.

To complete the picture, we also compare the quantum coherent time dependent dynamics with an approximation
based on the classical incoherent Wigner dynamics approximation to the quantum dynamics.
As also found using other
methods \cite{friedrich2}, the classical Wigner (incoherent) flight time associated with the reflection is found to be slightly {\it shorter} than
the quantum flight time as determined from the mean quantum transition path time.
This is but another indication that one should not think of the quantum particle as being reflected far away from
the surface, for if this were the case, the quantum flight time would have to be much shorter than that determined
from a classical Wigner approximation.
However, as might have been expected, at low energies, the quantum density is noticeable only far away from the surface while the
density generated through the Wigner dynamics does have appreciable amplitude close to the surface.

This nicely demonstrates Kohn's prediction \cite{kohn} that although classically the sticking coefficient at threshold
is unity, the quantum coefficient vanishes due to quantum reflection which prevents any noticeable amplitude of the
wavefunction close to the surface.
It is also consistent with the observation that He dimers do not dissociate upon quantum reflection \cite{zhao2}, since the density of the dimers is always sufficiently far from the surface to prevent any interaction which would break up the ultra weak bond.
We end in Section IV with a Discussion of the implications of the present study on quantum threshold surface scattering.

\renewcommand{\theequation}{2.\arabic{equation}} %
\setcounter{section}{1} \setcounter{equation}{0}

\section{Quantum reflection for general interaction potentials}

In 1988 Senn \cite{senn} established a quantum reflection theorem for one-dimensional scattering. He showed that the portion of particles that is transmitted in general vanishes as the kinetic energy of the incident
particles approaches zero. He also showed that this behavior is no longer valid when a bound level is present at the
onset of the continuum and in such cases, the reflection coefficient, even at threshold, is less than unity. In particular,
for symmetric potentials, the resonance condition implies that the reflection coefficient vanishes at threshold (threshold anomalies).
It is very instructive to briefly review the proof of  this important theorem since it lies at the heart of understanding
the quantum threshold reflection phenomenon.

Let us assume that the potential $V(x)$ differs appreciably  from zero only inside a finite interval
for which $-\xi < x < \xi$. Let $u$ and $v$ denote two linearly independent solutions of
the corresponding one-dimensional Schr\"odinger equation. For a particle incident from the left with positive momentum ($\hbar k$), the wavefunction
is
\begin{equation}
\Psi (x) = T e^{ikx}
\label{2.1}
\end{equation}
for $x \geq \xi$ and
\begin{equation}
\Psi (x) = e^{ikx} + R e^{-ikx}
\label{2.2}
\end{equation}
for $x \leq - \xi$, where $T$ and $R$ are the transmission
and reflection amplitudes, respectively.
Inside the region where the potential is different from zero, the wavefunction can be written as
\begin{equation}
\Psi (x) = a u(x) + b v(x)
\label{2.3}
\end{equation}
where the following boundary conditions, which assure independence of the two solutions, must be satisfied
\begin{equation}
v(-\xi) = u'(-\xi) = 0 , \, \, \, u(-\xi) = v'(- \xi) = 1
\label{2.4} .
\end{equation}
These boundary conditions imply that the Wronskian of $u$ and $v$ denoted by $W(u,v)=u v' - u' v$ is unity.

In addition to the boundary conditions one must impose continuity of the wavefunction  $\Psi$ and its first
derivative at $x=\pm \xi$:
\begin{eqnarray}
e^{ik\xi} + R e^{-ik\xi}& =& a \nonumber \\
ik(e^{ik\xi} - R e^{-ik\xi})& =& b \nonumber \\
T e^{ik\xi} &=& a u(\xi) + b v(\xi) \nonumber \\
ikT e^{ik\xi}& =& a u'(\xi) + b v'(\xi)  \label{2.5}  .
\end{eqnarray}
This set of linear equations is readily solved to express the reflection amplitude in terms of the various values
of the wavefunction and its first derivative:
\begin{equation}
R = e^{2ik \xi} \frac{k (p-q) + i(sk^2-w)}{k(p+q) + i(sk^2+w)}
\label{2.6}
\end{equation}
where $p=v'(\xi), q=u(\xi), s= -v(\xi)$ and $w=u'(\xi)$. Thus, the reflection probability
is given by
\begin{equation}
|R|^2 = \frac{w^2 + k^2[(p-q)^2 - 2 sw] + s^2 k^4}{w^2 + k^2[(p+q)^2 + 2 sw] + s^2 k^4} \label{2.7} .
\end{equation}
In the limit that  $k \rightarrow 0$ one finds that $|R|^2 \rightarrow 1$ unless $w$ tends
to zero as well. In this case
\begin{equation}
\label{threshold}
\lim_{k \rightarrow 0} |R|^2 = \frac{(p-q)^2}{(p+q)^2}
\end{equation}
and the reflection coefficient at threshold will in general differ from unity unless $p$ or $q$
vanish for $k \rightarrow 0$.
The condition $w=0$ in the limit $k \rightarrow 0$ holds if and only if the potential supports a
bound state at $E=0$.
This proof is given in detail in Ref.~\cite{senn}. For our purposes of studying quantum reflection, we will assume that $w\neq 0$.

From Eq.~(\ref{2.6}) we readily find that in the threshold limit
\begin{eqnarray}
\lim_{k\rightarrow 0}T(k)=-\frac{2ik}{w}, \ \ \lim_{k\rightarrow 0}R(k)=-\exp(2ik\xi)[1+\frac{2ikp}{w}]
\label{2.9}
\end{eqnarray}
implying that the transmission amplitude vanishes linearly with diminishing $k$ and the reflection amplitude goes to $-1$.

For surface scattering, that is in the case when the potential goes to infinity when $x$ becomes large,
the analysis simplifies,
since in this case the transmission coefficient vanishes, or in other words we have the  boundary condition that
the wavefunction vanishes when we go far enough to positive values of the coordinate.
The solution of Eqs.~(\ref{2.5}) simplifies to
\begin{eqnarray}
R(k)=\exp(2ik\xi)\frac{ks+iq}{ks-iq}
\label{2.10}
\end{eqnarray}
and quantum reflection takes the form:
\begin{eqnarray}
\lim_{k\rightarrow 0}R(k)=-\exp(2ik\xi)[1-\frac{2iks}{q}]
\label{2.11}.
\end{eqnarray}
Quantum reflection in this case may be identified by noting that the imaginary part of the reflection amplitude becomes linear with $k$.

To summarise, quantum reflection is a result only of the continuity of the wavefunction
and the boundary conditions.
It is a general phenomenon, unrelated to any classical feature such as badlands.
They are only meaningful when considering semiclassical approximations, but do not imply
that they cause the quantum reflection phenomenon.

\renewcommand{\theequation}{3.\arabic{equation}} %
\setcounter{section}{2} \setcounter{equation}{0}

\section{\protect\bigskip Quantum reflection for the Morse potential}

\subsection{Energy domain. Scattering wavefunctions}

In the previous section, an analysis of the behavior of the reflection coefficient for a general
one-dimensional potential, was reviewed.
The continuity requirement of the wavefunction as well and the boundary conditions are the only
conditions needed for quantum threshold reflection.
In this section, we present a detailed investigation of the threshold dynamics for the one-dimensional Morse potential.

The Schr\"odinger equation for the scattering of a particle of mass $m$ and incident energy
$E_i = \hbar^2 k^2 / 2 m$ (with initial momentum $\hbar k$)  by a Morse potential is
\begin{equation}
-\frac{\hbar ^{2}}{2m}\frac{d^{2}\Psi \left( z\right) }{dz^{2}}+V\left[ \exp
\left( -2\frac{z-z_{0}}{d}\right) -2\exp \left( -\frac{z-z_{0}}{d}\right) %
\right] \Psi \left( z\right) =\frac{\hbar ^{2}k^{2}}{2m}\Psi \left( z\right)
\label{3.1}
\end{equation}%
%
where $V$ is the well depth, $d^{-1}$ is the stiffness parameter, the minimum of the potential is located at
$z=z_{0}$ and the harmonic frequency of motion about the well bottom is
\begin{equation}
\omega _{0}^{2}=\frac{2V}{md^{2}}. \label{3.2}
\end{equation}%
(We have chosen here to have the particle incident from the right rather than from the left as in the previous section.) Introducing the dimensionless ``coordinate''  \cite{morse,matsumoto}%
\begin{equation}
y=Y\exp \left( -\frac{z-z_{0}}{d}\right)
\label{3.3}
\end{equation}%
and the reduced variables:
\begin{equation}
\mu ^{2}=-d^{2}k^{2},1=\frac{8md^{2}V}{\hbar ^{2}Y^{2}}
\label{3.4}
\end{equation}%
where the parameter $Y$ is expressed in terms of the other parameters, allows us to rewrite the Schr{\"o}dinger equation \ref{3.1} as:%
\begin{equation}
\frac{1}{y}\frac{d\Psi \left( y\right) }{dy}+\frac{d^{2}\Psi \left( y\right)
}{dy^{2}}+\left[ -\frac{1}{4}+\frac{Y}{2y}-\frac{\mu ^{2}}{y^{2}}\right]
\Psi \left( y\right) =0  .\label{3.5}
\end{equation}
Using the substitution %
\begin{equation}
\Psi \left( y\right) =y^{\mu }\exp \left( -\frac{y}{2}\right) u\left(
y\right) \label{3.6}
\end{equation}%
leads to the differential equation for the confluent hypergeometric function \cite{abramo}
\begin{equation}
y\frac{d^{2}u\left( y\right) }{dy^{2}}+\left( 1+2\mu -y\right) \frac{%
du\left( y\right) }{dy}-\left( \frac{1+2\mu -Y}{2}\right) u\left( y\right) =0.\label{3.7}
\end{equation}
Its two independent solutions $u$ and $v$ may be chosen to be%
\begin{equation}
u\left( y\right) =M\left( \frac{1+2\mu -Y}{2},1+2\mu ,y\right) ,v\left(
y\right) =y^{-2\mu }M\left( \frac{1-2\mu -Y}{2},1-2\mu ,y\right)
\label{3.8}
\end{equation}%
where $M$ is Kummer's function as defined in Ref. \cite{abramo}.

The scattering wavefunctions are a linear combination of the two independent solutions:
\begin{equation}
\Psi^+(y)=y^{\mu}\exp \left( -\frac{y}{2}\right)\left[A u\left(
y\right)+Bv\left(y\right)\right]\label{3.9}.
\end{equation}
To determine the coefficients one imposes the boundary conditions.
When $z \rightarrow \infty $
with $k>0$, the scattering wavefunction has the form%
\begin{equation}
\Psi^+ \left( z\right) =\frac{1}{\sqrt{2\pi }}\left[ \exp \left( -ikz\right)
+R\left( k\right) \exp \left( ikz\right) \right] \label{3.10}
\end{equation}
and $R(k)$ is the reflection amplitude.
The second boundary condition is that the wavefunction vanishes in the limit $z\rightarrow -\infty $.
These boundary conditions are another indication that the full potential region is needed to correctly extract
the reflection amplitude, not only its long tail.

Noting the asymptotic property of the Kummer function
\begin{equation}
\lim_{y\rightarrow 0}M\left( \frac{1+2\mu -Y}{2},1+2\mu ,y\right) =1 \label{3.11}
\end{equation}%
we find that when $z\rightarrow \infty $%
\begin{eqnarray}
\lim_{z\rightarrow \infty }\Psi^+ \left( z\right) &=&Ay^{\mu }+By^{-\mu }  \nonumber \\
&=&AY^{ikd}\exp \left( ikz_{0}\right) \left[ \exp \left( -ik\left(
z-z_{0}\right) \right) +\frac{B}{A}Y^{-2ikd}\exp \left( ikz\right) \exp
\left( -2ikz_{0}\right) \right] \nonumber \\ \label{3.12}
\end{eqnarray}%
from which we identify the reflection amplitude
\begin{equation}
R\left( k\right) =\frac{B}{A}\exp \left( -2ikz_{0}\right) Y^{-2ikd} \label{3.13}
\end{equation}%
and its modulus or reflectivity as $|R| = B/A$.

The second boundary condition is for $z\rightarrow -\infty $ or equivalently $%
y\rightarrow \infty $. Noting the properties %
\begin{eqnarray}
\lim_{y\rightarrow \infty }M\left( \frac{1+2\mu -Y}{2},1+2\mu ,y\right) &=&%
\frac{\Gamma \left( 1+2\mu \right) }{\Gamma \left( \frac{1+2\mu -Y}{2}%
\right) }\exp \left( y\right) y^{\frac{-1-2\mu -Y}{2}}\label{3.14} \\
\lim_{y\rightarrow \infty }y^{-2\mu }M\left( \frac{1-2\mu -Y}{2},1-2\mu
,y\right) &=&\frac{\Gamma \left( 1-2\mu \right) }{\Gamma \left( \frac{1-2\mu
-Y}{2}\right) }\exp \left( y\right) y^{\frac{-1-2\mu -Y}{2}}\label{3.15}
\end{eqnarray}%
and imposing the boundary condition that the wavefunction vanishes when $z\rightarrow-\infty$ implies that:
\begin{equation}
|R(k)| = \frac{B}{A}=-\frac{\Gamma \left( 1+2\mu \right) \Gamma \left( \frac{1-2\mu -Y%
}{2}\right) }{\Gamma \left( 1-2\mu \right) \Gamma \left( \frac{1+2\mu -Y}{2}%
\right) } \label{3.16}
\end{equation}%
and the reflection amplitude is given by%
\begin{equation}
R\left( k\right) =-\frac{\Gamma \left( 1+2\mu \right) \Gamma \left( \frac{%
1-2\mu -Y}{2}\right) }{\Gamma \left( 1-2\mu \right) \Gamma \left( \frac{%
1+2\mu -Y}{2}\right) }\exp \left( -2ikz_{0}\right) Y^{-2ikd}
\label{3.17}
\end{equation}%
From these results one notes that as expected $R^*(k) = R(-k)$ and $\left\vert R\left( k\right) \right\vert ^{2} = 1$. %
In the context of quantum threshold reflection we readily establish that in this limit:%
\begin{equation}
\lim_{k\rightarrow 0}R\left( k\right)\exp(2ikz_0)
=-1+2idk\left[\ln(Y)+d\bar{\Psi}\left(\frac{1+Y}{2}\right)-\pi\tan\left(\frac{\pi Y}{2}\right)+4\gamma\right]
\label{3.18}
\end{equation}%
where $\bar{\Psi}$ is the digamma function and $\gamma=0.5772...$ is Euler's constant.



\begin{figure}
\includegraphics[scale=1]{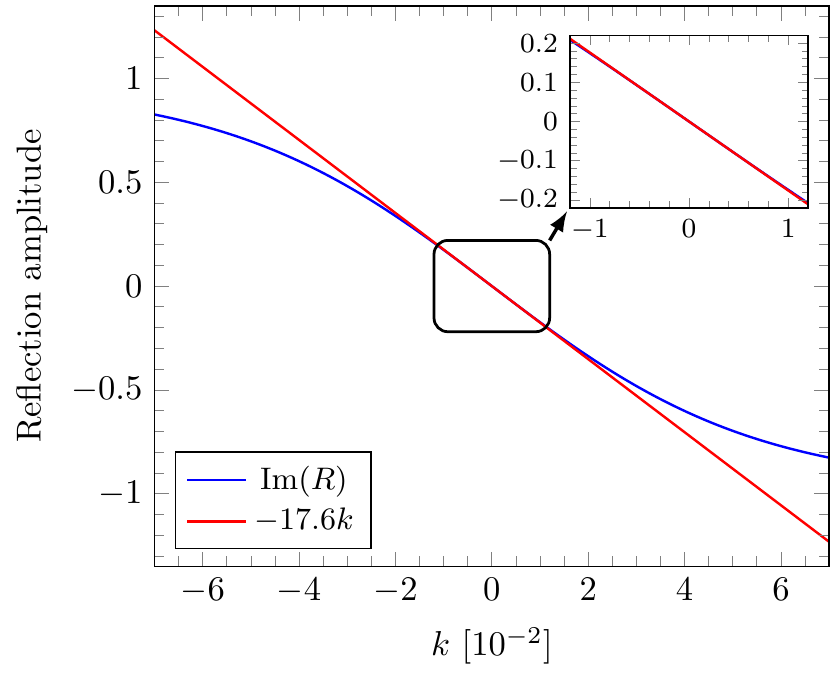}
\caption{Imaginary part of the reflection amplitude as a function of $k$. The region around
$k=0$ is enhanced.}
\label{reflection}
\end{figure}

To exemplify the quantum reflection threshold region we plot in Fig.~\ref{reflection} the
imaginary part of the reflection amplitude as a function of $k$.
For this purpose, and throughout this paper, we use atomic units, with $\hbar=m=1$ and
the parameters of the Morse potential are taken to be $V=1, d=1, z_0 = 0$ so that $\omega_0 = \sqrt{2}$ and
$Y=2 \sqrt{2}$.
We note the linear dependence of the imaginary part of the amplitude about $k=0$.
The numerical slope is $-17.626$ (see also Eq.~(\ref{3.18})).
The interval of linearity is roughly $\left[-10^{-2}\leq k\leq 10^{-2}\right]$ and
this is the domain of quantum threshold reflection for the Morse potential (using the parameters as above).

This range of $k$ values is also the region in which the absolute value of
the badlands function for the Morse potential becomes greater than unity.
Denoting the classical momentum as
\begin{equation}
p(z)=\pm\sqrt{(2m[E-V(z)])}
\label{3.19}
\end{equation}
the badlands function is defined as \cite{friedrich13}:
\begin{equation}
Q(z)=\hbar^2\left(\frac{3}{4}\frac{(p')^2}{p^4}-\frac{p''}{2p^3}\right)
\label{3.20}
\end{equation}
where primes denote derivatives with respect to the argument.
In Fig.~2 we plot the absolute value of the badlands function vs.\ the
coordinate $z$ in the range of $k$ values for which the scattering is dominated by quantum threshold reflection.
As expected, the Morse potential exhibits a badlands region, the location of the
regions in which $|Q(z)|>1$ moves outward and the magnitude of the function increases with decreasing $k$.
\begin{figure}
\includegraphics[scale=1]{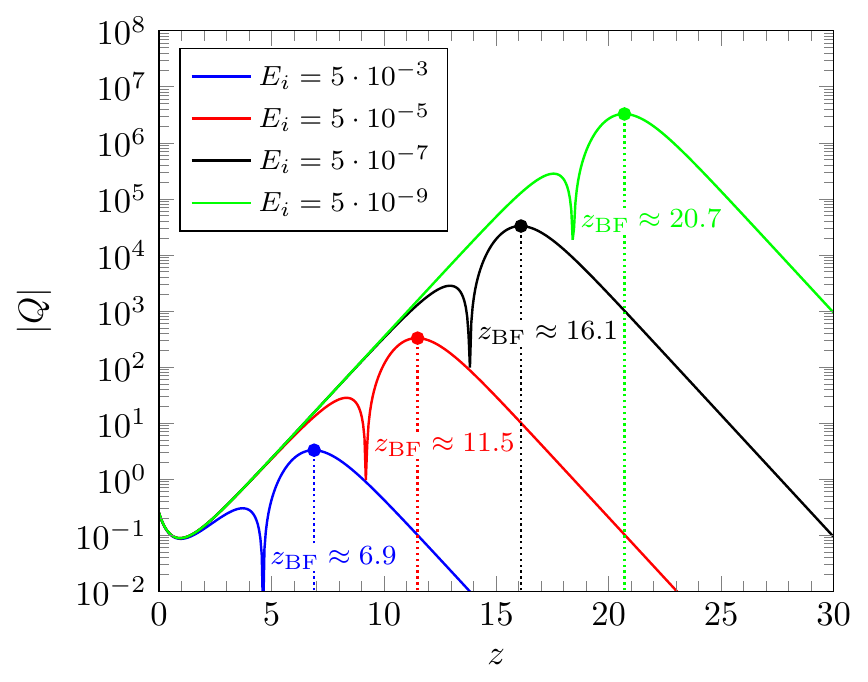}
\caption{Absolute value of the badlands function $|Q|$ as a function of the distance $z$ for various incident energies.
The badlands region, is localized around the value $z_\text{BF}$ where $|Q|$ attains its maximal value. }
\label{badlands}
\end{figure}
%


Imposing the boundary condition for $z\rightarrow\infty$ allows us to identify that
\begin{equation}
A\exp \left( ikz_{0}\right) Y^{ikd}=\frac{1}{\sqrt{2\pi }} \label{3.21}
\end{equation}%
so that the final expression for the scattering wavefunction for a given value of the
incident momentum $\hbar k$ is:%
\begin{eqnarray}
&&\langle x\vert k^+\rangle\equiv \Psi^+ _{k}\left( z\right) =\frac{1}{\sqrt{2\pi }}\exp \left( -\frac{y}{2}%
\right) \nonumber \\
&&\cdot \left[ \exp \left( -ikz\right) M\left( \frac{1+2ikd-Y}{2},1+2\mu
,y\right)
+R\left( k\right) \exp \left( ikz\right) M\left( \frac{1-2ikd-Y}{2}%
,1-2ikd,y\right) \right] \nonumber \\ \label{3.22}
\end{eqnarray}%
This result will be used in the next subsection to construct the propagator for the Morse potential.

\subsection{Time domain. Wave packet propagation}

For the scattering from the Morse potential, there is no transmission so that the completeness of the
scattering states is expressed as
\begin{equation}
\hat{I}=\int_{0}^{\infty }dk\left\vert k^{+}\rangle \langle k^{+}\right\vert \label{3.23}  .
\end{equation}%
The coordinate space matrix element of the propagator is then:
\begin{eqnarray}
\left\langle z\left\vert \exp \left( -\frac{i\hat{H}t}{\hbar }\right)
\right\vert z' \right\rangle  =\int_{0}^{\infty }dk\exp \left( -i\frac{\hbar k^{2}}{2m}t\right) \langle
z\left\vert k^{+}\rangle \langle k^{+}\right\vert z' \rangle .
\label{3.24}
\end{eqnarray}%

We will study the spatial and temporal dynamics of an initial coherent state centered about the
position $z_i$ and incident momentum $-p_i$ ($p_i>0$) and characterized by the width parameter $\Gamma $.
Its coordinate representation is:%
\begin{equation}
\label{wavepacket}
\left\langle z|\Phi \right\rangle =\left( \frac{\Gamma }{\pi }\right)
^{1/4}\exp \left( -\Gamma \frac{\left( z-z_{i}\right) ^{2}}{2}+\frac{i}{%
\hbar }p_{i}\left( z_{i}-z\right) \right) .
\end{equation}%
Since the incident wavefunction is localized in the asymptotic region, we
may readily evaluate the matrix element $\left\langle k^{+}|\Phi
\right\rangle $ using the asymptotic form of the wavefunction:%
\begin{eqnarray}
\left\langle k^{+}|\Phi \right\rangle  &=&
\left( \frac{1}{\pi \Gamma }\right) ^{1/4}\left[ \exp \left(ikz_{i} -\frac{1}{2\Gamma }\left( \frac{p_{i}}{\hbar }%
-k\right) ^{2}\right) +R^{\ast }\left( k\right)
\exp \left( -ikz_{i}-\frac{1}{2\Gamma }\left( \frac{p_{i}}{\hbar }+k\right)
^{2}\right) \right] \nonumber \\ \label{3.26}
\end{eqnarray}%

The time evolution of the incident wavepacket is then given by
\begin{eqnarray}
&&\left\langle z\left\vert \exp \left( -\frac{i\hat{H}t}{\hbar }\right)
\right\vert \Phi \right\rangle  =\int_{0}^{\infty }dk\exp \left( -i\frac{\hbar k^{2}}{2m}t\right) \langle
z\left\vert k^{+}\rangle \langle k^{+}\right\vert \Phi \rangle \notag \\
&&=\left( \frac{1}{\pi \Gamma }\right) ^{1/4}\frac{1}{\sqrt{2\pi }}\exp
\left( -\frac{Y}{2}\exp \left( -\frac{z-z_{0}}{d}\right) \right)  \nonumber \\
&& \left[ \int_{-\infty }^{\infty }dk\exp \left( -i\frac{\hbar k^{2}}{2m}t-%
\frac{1}{2\Gamma }\left( \frac{p_{i}}{\hbar }-k\right) ^{2}-ik\left(
z-z_{i}\right) \right) M\left( \frac{1+2ikd-Y}{2},1+2ikd,y\left( z\right)
\right) \right.  \nonumber \\
&& \left. +\int_{-\infty }^{\infty }dk\exp \left( -i\frac{\hbar k^{2}}{2m}t-%
\frac{1}{2\Gamma }\left( \frac{p_{i}}{\hbar }-k\right) ^{2}+ik\left(
z+z_{i}\right) \right) R\left( k\right) M\left( \frac{1-2ikd-Y}{2}%
,1-2ikd,y\left( z\right) \right) \right]  \nonumber \\ \label{propagation}
\end{eqnarray}%
where we used the fact that $R^{\ast }\left( k\right)
=R\left( -k\right)$. %
With these preliminaries, the time propagation is reduced to a numerical quadrature over $k$.
In practice, the propagation of this initial wave packet expressed in Eq.~(\ref{propagation}) is
carried out by integration over $k$ in the range
$\left[p_{i}/\hbar-7\sqrt{\Gamma};p_{i}/\hbar+7\sqrt{\Gamma}\right]$. The overlap $|\langle k^+ | \Phi\rangle|$
is virtually zero outside this range.
The number of $k$-grid points is $5\cdot10^{4}$ and this leads to converged evaluation of the integral
in Eq.~(\ref{propagation}) for any combination of $E_i$, $\Gamma$, $t$, and $z$ used in this study.

\begin{figure}
\includegraphics[scale=1]{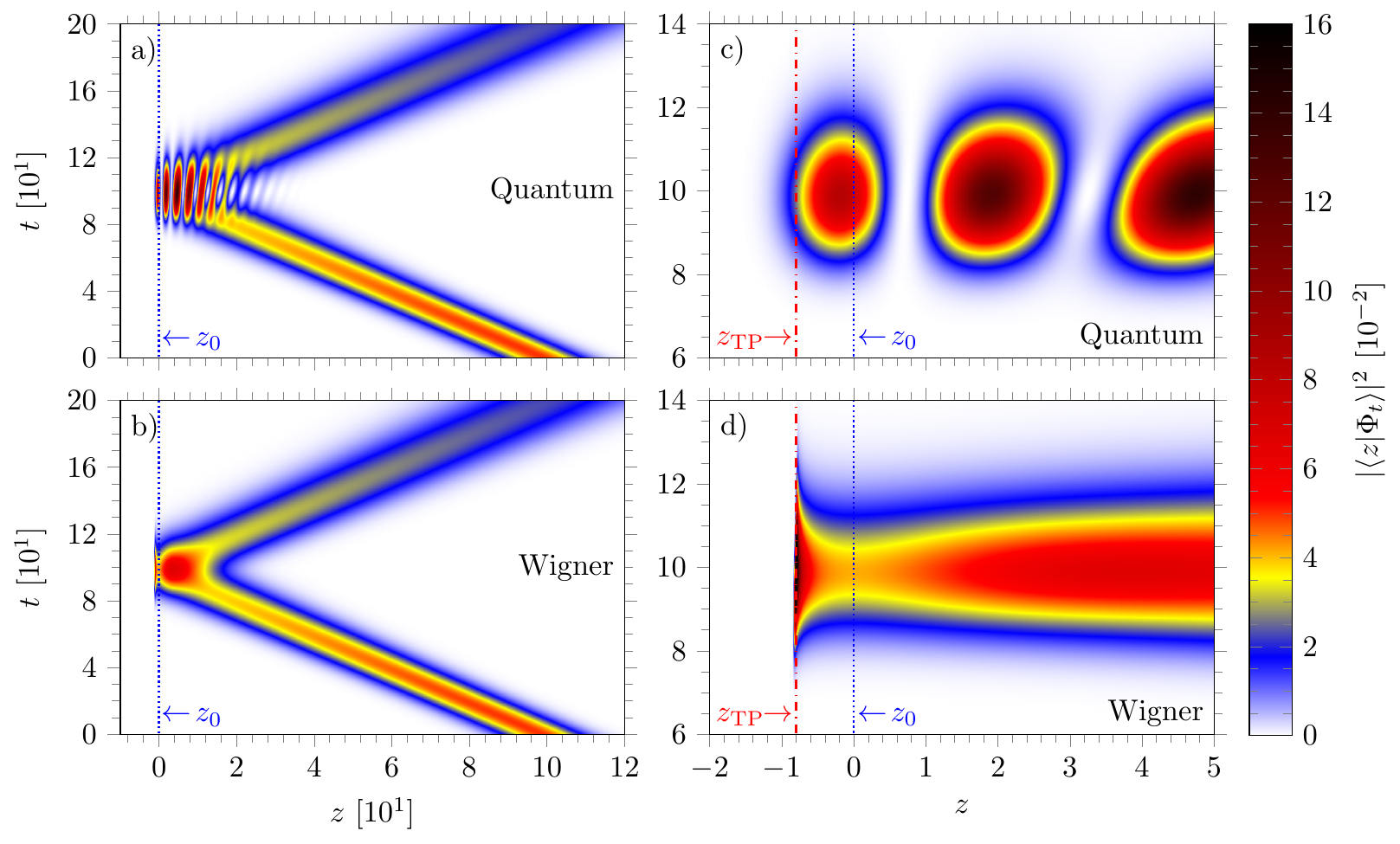}
\caption{
The space and time evolution of a coherent state scattered from a Morse potential at the incident energy $E_i=0.5$. Panels a) and b) show the time and space evolution of the probability density using exact quantum and classical Wigner propagation, respectively.
Panels c) and d) show a zoom of the same time evolutions close to the potential well.
The minimum of the Morse potential $z_0$ and the classical turning point $z_\text{TP}$ (corresponding to $E_i$) are indicated.
At this relatively high energy the quantum wavepacket approaches the repulsive wall and even
tunnels beyond it.
For further details see the text.
}
\label{propagation1}
\end{figure}
\begin{figure}
\includegraphics[scale=1]{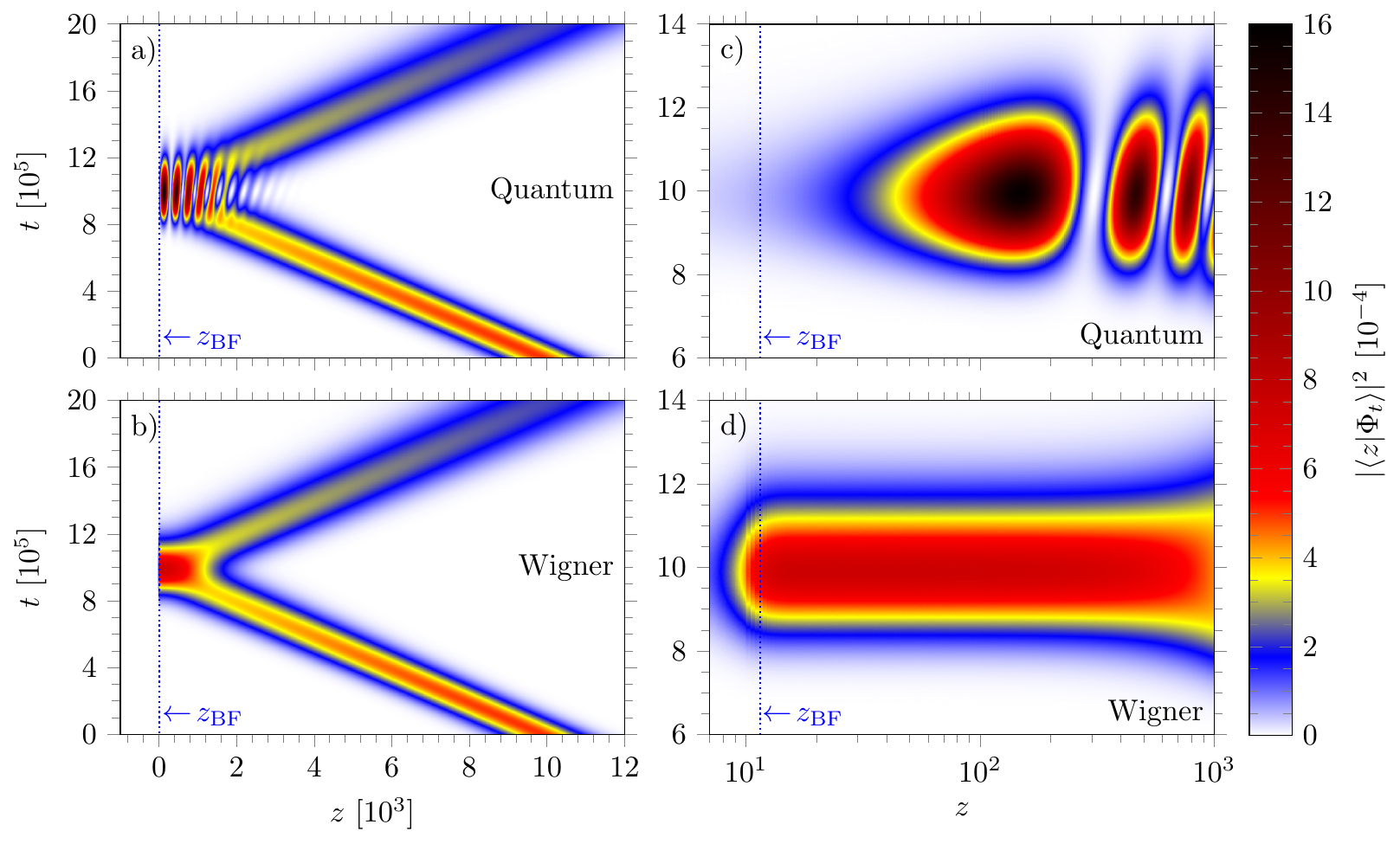}
\caption{
The space and time evolution of a coherent state scattered from a Morse potential at the incident energy $E_i=5\cdot 10^{-5}$. Panels a) and b) show the time and space evolution of the probability density using exact quantum and classical Wigner propagation, respectively.
Panels c) and d) show a zoom of the same time evolutions close to the potential well.
The maximum of the absolute value of the badlands function $|Q(z)|$ at this incident energy is at $z_\text{BF}\approx 11.5$.
}
\label{propagation2}
\end{figure}
\begin{figure}
\includegraphics[scale=1]{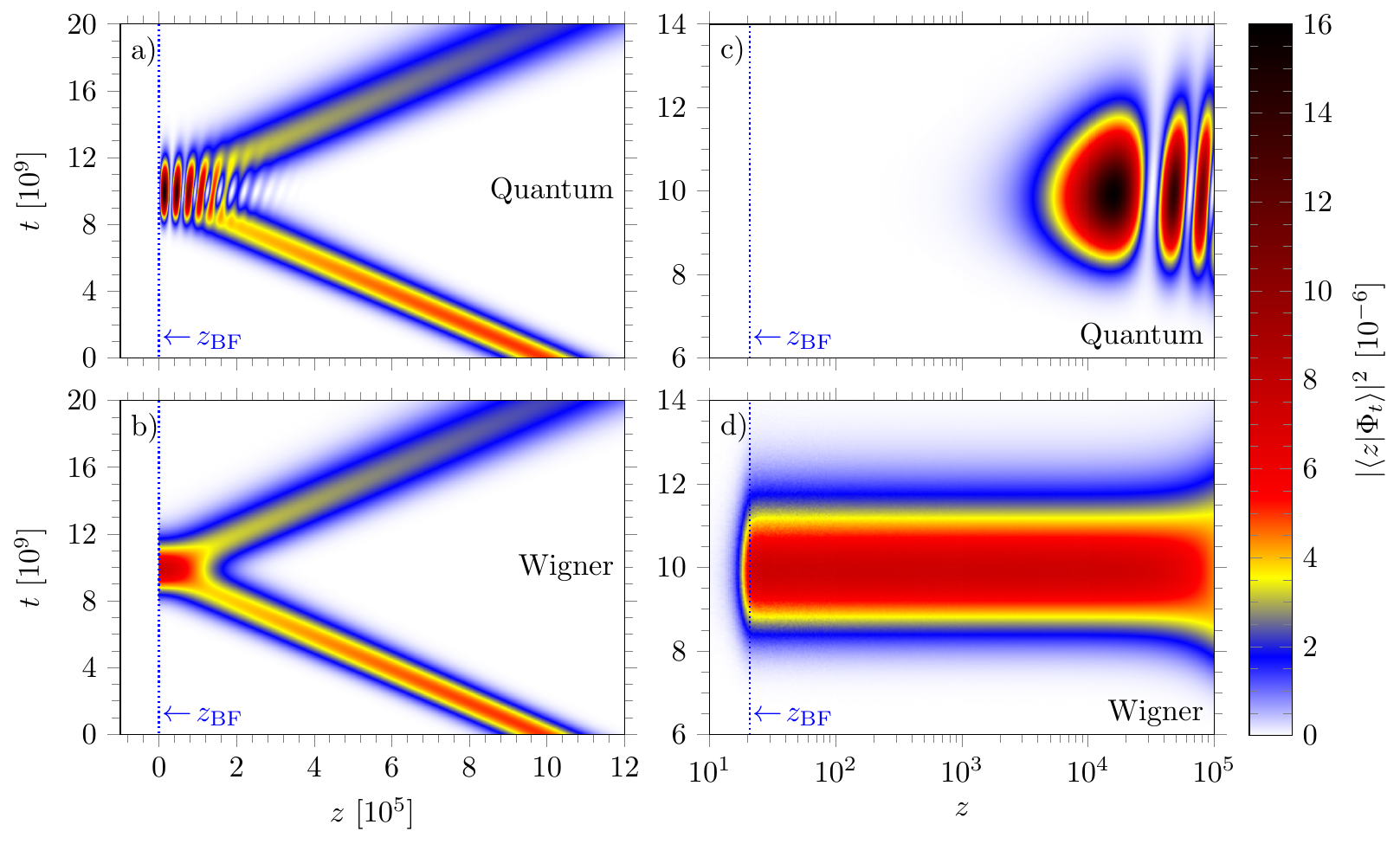}
\caption{
The space and time evolution of a coherent state scattered from a Morse potential at the incident energy $E_i=5\cdot 10^{-9}$. Panels a) and b) show the time and space evolution of the probability density using exact quantum and classical Wigner propagation, respectively.
Panels c) and d) show a zoom of the same time evolutions close to the potential well.
The maximum of the absolute value of the badlands function $|Q(z)|$ at this incident energy is at $z_\text{BF}\approx 20.7$.
}
\label{propagation3}
\end{figure}

The time evolution of the initial wave packet is plotted in Figs.~\ref{propagation1},
\ref{propagation2} and \ref{propagation3} for three different incident energies
($E_i=p_i^2/2$): $5\cdot 10^{-1}$, $5\cdot 10^{-5}$, and $5\cdot 10^{-9}$, respectively.
The initial parameters for the three energies are $(-p_i,z_i)=(-1,10^2)$ and $\Gamma = 10^{-2}$,
$(-p_i,z_i)= (-10^{-2},10^4)$ and $\Gamma = 10^{-6}$, and $(-p_i,z_i)=(-10^{-4},10^6)$ with
$\Gamma = 10^{-10}$, respectively.
Four panels are displayed in each figure.
Panel a) shows the quantum time evolution of the probability density of the coherent state evaluated from Eq.~(\ref{propagation}).
Panel b) shows the corresponding time evolution of $|\langle z | \Phi \rangle|^2$ based on a classical
Wigner dynamics approximation to the exact quantum dynamics.
In the classical Wigner propagation, classical trajectories are launched with initial conditions
sampled from the Wigner transform of $\langle z | \Phi \rangle$ (in all cases $4\cdot 10^9$ trajectories
are included), and the relative probability density at $z^\prime$ and $t^\prime$ is calculated by
counting the trajectories with position $z^\prime$ at time $t^\prime$.
Panels c) and d) show a blow-up of the time-dependent probability density close to the well of the
Morse potential for the exact quantum and classical Wigner dynamics, respectively.
In all panels, the number of grid points in both $z$- and $t$-space is 2000.

In Fig.~\ref{propagation1} the dotted blue line (denoted by $z_0$) shows the location of the Morse
potential well, and the dash-dotted red line (denoted by $z_\text{TP}$) shows the location
of the classical turning point at the incident momentum $p_i$.
In this relatively high energy case, the magnitude of the badlands function remains small so that
in any case it is not important.
In Figs.~\ref{propagation2} and \ref{propagation3} the dotted blue line (denoted by $z_\text{BF}$) shows
the location of the maximal magnitude of absolute value of the badlands function.

As shown in Fig.~\ref{propagation1} at the highest energy probed ($E_i=0.5$), which is at the edge of
the region in which the imaginary part of the reflection amplitude is still linear in $k$
(see Fig.~\ref{reflection}), the quantum wavepacket reaches the left turning point of the Morse potential
and even penetrates into the classically forbidden region.
For the classical Wigner approximation, the probability density builds up at the classical turning point
$z_\text{TP}$ (with a slight ``penetration'' due to the momentum spread around $p_i$ of the Wigner transform
of $\langle z|\Phi\rangle$) and has a relatively small amplitude at the bottom of the potential well, where
the classical trajectories spend less time. Note that even at this relatively high energy, there is a qualitative difference between the quantum and classical Wigner scattering. The former shows an oscillatory structure typical for the interference of incoming and outgoing waves, while the incoherent Wigner distribution does not show this at all.

At the two lower incident energies shown in Figs. \ref{propagation2} and \ref{propagation3}, the quantum wavepacket does not reach the interaction region and the reflection takes place at distances that are far greater than $z_\text{BF}$, where
the maximum of the absolute value of the badlands function is found.
From these two plots it becomes evident that the badlands region has no special meaning for
the quantum evolution.
Especially at the lowest energy probed (Fig.~\ref{propagation3}) the first maximal density is
located at a distance which is $\sim$$1000$ times larger than the location of the maximum of the badlands
function ($z_\text{BF}$).
No special attention should then be paid to the badlands function.
The repulsive wall of the potential or its left turning point also plays no role in this
coherent interference process except for imposing the boundary condition that the function vanishes
to the left of the repulsive wall.
On the other hand, the classical Wigner approximation of the wavepacket dynamics leads to a significant
probability density even at the badlands region.
This is not surprising since the badlands region is positioned in the tail of the Morse potential, where
the potential is almost constant.
Hence, the time spent by the classical trajectories in this region is not different from the remaining part
of the potential tail.


As may also be inferred from Figs. \ref{propagation2} and \ref{propagation3} the quantum reflection is just a result of the coherent interference of the incoming and outgoing wavefunctions. Since the particle is almost a free particle, the boundary condition that the wavefunction vanishes a bit further left to the classical turning point, implies that to a good approximation the wavefunction is just $\sim \sin[k(z-z_{\text{TP}})]$ which is just the difference between the amplitude of the incoming and outgoing waves. Inspection of the two figures shows that at the respective inflection times $t=10^6,10^{10}$ the maximum of the wavefunction in the two respective figures occurs at $z-z_{\text{TP}}\simeq \pi/(2k_i)$ while the first zero occurs at $z-z_{\text{TP}}\simeq \pi/k_i$. Changing the location of the turning point, for example, by making the Morse potential softer, would change the location of the respective maxima. This is again another indication of the
nonlocal character of this scattering or coherent interference process.

\subsection{Mean flight time}

In this subsection we will compare the mean flight time of the initial wave packet
for arrival at a position $y=2z_i$ as derived from the exact quantum mechanics and a classical Wigner approximation. For this purpose, we note that the transition path time distribution \cite{hummer04}, in a quantum
mechanical context is defined in terms of the positive density correlation function at time $t$ about the final position $y$
\begin{equation}
C_{t}\left(y;\Phi\right)=\text{Tr}\left[\left|\Phi\left\rangle \right\langle \Phi\right|\hat{K}_{t}^{\dagger}
\delta\left(\hat{z}-y\right)\hat{K}_{t}\right]=\left|\left\langle y\left|\hat{K}_{t}\right|\Phi\right\rangle \right|^{2}, \label{eq:ct}
\end{equation}
where $\hat{K}_{t}=\exp\left(-i\hat{H}t/\hbar\right)$ is the quantum propagator \cite{pollak17a,pollak17b,pollak17c,petersen17}.
The transition path time probability distribution reads
\begin{equation}
P_{t}\left(y;\Phi\right)=\frac{C_{t}\left(y;\Phi\right)}{\int_{0}^{\infty}\text{d}t\, C_{t}\left(y;\Phi\right)},
\end{equation}
where we assume that the normalization integral ${\int_{0}^{\infty}\text{d}t\, C_{t}\left(y;\Phi\right)}$ is finite. The mean flight time for arrival at $y$ is then by definition
\begin{equation}
\left\langle t\right\rangle_\text{QM} =\int_{0}^{\infty}dt\, tP_{t}(y;\Phi).
\end{equation}

The correlation function of Eq.~(\ref{eq:ct}) can be rewritten as a phase space trace of two Wigner
densities.
One is the Wigner representation of $\langle z|\Phi \rangle$ and the other is the Wigner
representation of the Heisenberg time evolved density $\hat{\rho}\left( t\right) =\exp\left(i\hat{H}t\right) \delta
\left( \hat{z}-y\right) \exp \left( -i\hat{H}t\right)$.
The classical Wigner approximation is then obtained by replacing the exact Wigner representation of the time evolved quantum density with its classical
Wigner approximation $\delta \left( q_{t}-y\right)$ where $q_{t}$ is the classical trajectory that is
evolved to time $t$ from the initial condition $(p,q)$. This enables us to calculate the mean flight time based on the classical Wigner approximation, which we denote
$\langle t \rangle_\text{W}$.

The resulting mean flight times as a function of the incident energy are shown in Table~\ref{table1}.
First we notice that the mean flight times based on the classical Wigner approximation are
consistently smaller than the corresponding times based on exact quantum propagation.
Naively, this might not be expected from visual inspection of Figs. \ref{propagation2} and \ref{propagation3}, which show that the quantum
probability density in the neighbourhood of the potential well is much smaller than the Wigner
probability density. In different words, if the quantum reflection phenomenon would be a result of reflection from the badlands region, the mean flight time of the quantum particle should be {\it less} than that of the classical Wigner flight time since the path traversed is shorter.
The fact that $\langle t\rangle_\text{W} < \langle t\rangle_\text{QM}$ indicates that the quantum
particle is not being reflected far away from the potential well.
Furthermore, the mean flight time of the free (classical) particle for traversal of the distance
$y+z_{i}+2\left|z_{\text{TP}}\right|$, where $z_\text{TP}$ is the classical turning point corresponding to
$E_i$, is consistently shorter \cite{friedrich2} than either  $\langle t \rangle_\text{QM}$ or $\langle t \rangle_\text{W}$.
Since the classical Wigner approximation is exact for free particle motion and $\langle t\rangle_\text{free}<\langle t\rangle_\text{W}$, another indication that one should not associate a physical turning point to the quantum reflection phenomenon.

\begin{table}
\small
\centering
\caption{
Mean flight time associated with quantum threshold reflection. $E_{i}$ is the incident energy, $\left(-p_{i},z_{i}\right)$
is the phase space center of $\langle z |\Phi\rangle$, and $\Gamma$
is its width parameter.
$z_\text{TP}$ is the classical turning point corresponding to $E_i$.
$\left\langle t\right\rangle _{\text{free}}$ is the mean flight time
for a free (classical) particle to traverse the distance $y+z_{i}+2\left|z_{\text{TP}}\right|$.
$\left\langle t\right\rangle _{\text{QM}}$ is the mean flight time
for arrival at $y$ based on exact propagation of $\langle z|\Phi\rangle$.
$\left\langle t\right\rangle _{\text{W}}$ is the mean flight time for arrival
at $y$ based on classical Wigner propagation of $\langle z|\Phi\rangle$.
}
\begin{tabular}{ccccccc}
\noalign{\vskip\doublerulesep}
$E_i$ & $\left( -p_i,z_i \right)$ & $\Gamma$ & $z_\text{TP}$ & $\langle t\rangle_\text{free}$ & $\langle t\rangle_\text{QM}$ & $\langle t\rangle_\text{W}$ \tabularnewline[\doublerulesep]
\hline
\noalign{\vskip\doublerulesep}
\noalign{\vskip\doublerulesep}
$5\cdot10^{-1}$ & $\left(-1,10^2\right)$       & $10^{-2}$  & $-0.7996422$ & $3.0159928\cdot10^2$    & $3.0271643\cdot10^2$    & $3.0247033\cdot10^2$ \tabularnewline[\doublerulesep]
\noalign{\vskip\doublerulesep}
\noalign{\vskip\doublerulesep}
$5\cdot10^{-3}$ & $\left(-10^{-1},10^3\right)$ & $10^{-4}$  & $-0.6942948$ & $3.0013888\cdot10^4$    & $3.0353657\cdot10^4$    & $3.0199008\cdot10^4$ \tabularnewline[\doublerulesep]
\noalign{\vskip\doublerulesep}
\noalign{\vskip\doublerulesep}
$5\cdot10^{-5}$ & $\left(-10^{-2},10^4\right)$ & $10^{-6}$  & $-0.6931597$ & $3.0001386\cdot10^6$    & $3.0325289\cdot10^6$    & $3.0273962\cdot10^6$ \tabularnewline[\doublerulesep]
\noalign{\vskip\doublerulesep}
\noalign{\vskip\doublerulesep}
$5\cdot10^{-7}$ & $\left(-10^{-3},10^5\right)$ & $10^{-8}$  & $-0.6931473$ & $3.0000139\cdot10^8$    & $3.0309571\cdot10^8$    & $3.0289827\cdot10^8$ \tabularnewline[\doublerulesep]
\noalign{\vskip\doublerulesep}
\noalign{\vskip\doublerulesep}
$5\cdot10^{-9}$ & $\left(-10^{-4},10^6\right)$ & $10^{-10}$ & $-0.6931472$ & $3.0000014\cdot10^{10}$ & $3.0307969\cdot10^{10}$ & $3.0292251\cdot10^{10}$ \tabularnewline[\doublerulesep]
\noalign{\vskip\doublerulesep}
\end{tabular}
\label{table1}
\end{table}

\section{Concluding remarks}

In this work, we have clearly shown that there is no need to invoke the nonclassical or badlands
region of the interaction potential to understand the phenomenon of quantum threshold reflection. At the threshold energies, imposing the condition that the wavefunction vanishes beyond the classical turning point of the potential implies that the scattering wavefunction is well approximated as $\sin(k(z-z_{\text{TP}}))$. Since the incident wavevector is very small, the first maximum of the wavefunction occurs far away from the turning point. As a result the density around the potential well is very small and any interaction with the surface is negligible. Comparison with results based on a classical Wigner approximation serve to stress that quantum reflection is a result of the coherent sum of an incoming and an outgoing wave and it is the destructive and constructive interference that combines to give a density which is very small in the strong interaction region. This is why for example, the He dimer does not dissociate upon scattering from a surface at very low energies. This has nothing to do with the badlands region.

Quantum threshold reflection is thus a nonlocal interference process, in which the incoming and outgoing waves interfere destructively in the region of strong interaction, due to the boundary condition that the wavefunction must vanish beyond the classical turning point region. The badlands region has implications when attempting to construct a WKB solution to the scattering wavefunction, but it has no implication on the quantum dynamics.
This study should serve to definitely rule out the badlands region as a criterion for quantum threshold reflection.


\newpage

\noindent
{\bf Acknowledgement}
This work was supported by a grant from the
Israel Science Foundation and was partially supported by a grant with Ref.
FIS2014-52172-C2-1-P from the Ministerio de Economia y Competitividad
(Spain).

\end{document}